\documentstyle{article}

\newlength{\defaultparindent}
\newenvironment{Default Paragraph Font}{}{}
\input{tcilatex}

\begin{document}

{\large Compatibility of the Expansive Nondecelerative Universe Model with}

{\large the Newton Gravitational Theory and the General Theory of
Relativity}

\thinspace

{\bf Miroslav Sukenik}$^{a}${\bf , Jozef Sima}$^{a}${\bf \ and Julius
Vanko}$%
^{b}$

$^{b}${\small Slovak Technical University, Dep. Inorg. Chem., Radlinskeho 9,
812 37 Bratislava, Slovakia}

$^{a}${\small Comenius University, Dep. Nucl. Physics, Mlynska dolina F1,
842 48 Bratislava, Slovakia}

{\small e-mail: }{\it sima@chelin.chtf.stuba.sk; vanko@fmph.uniba.sk}

{\bf Abstract.} Applying the Vaidya metrics in the model of Expansive
Nondecelerative Universe (ENU) leads to compatibility of the ENU model both
with the classic Newton gravitational theory and the general theory of
relativity in weak fields.

Applying the Vaidya metrics [1] to gravitational field has led us [2] to
relation

$\varepsilon _{g}=-\frac{R.c^{4}}{8\pi .G}=-\frac{3m.c^{2}}{4\pi .a.r^{2}}%
\qquad \qquad \qquad \qquad \qquad \qquad \ $(1)\

where $\varepsilon _{g}$ is the density of the gravitational energy created
by a body with the mass {\it m} in the distance {\it r}, {\it R} is the
scalar curvature, {\it a} is the gauge factor reaching at the present

$a\approx 10^{26}m\qquad \qquad \qquad \qquad \qquad \qquad \qquad \qquad
\,\ \ \ \ \ $(2)

The mean energy density of ENU, $\varepsilon _{ENU}$ is expressed [3] as

$\varepsilon _{ENU}=\frac{3c^{4}}{8\pi .G.a^{2}}\qquad \qquad \qquad \qquad
\qquad \qquad \qquad \ \ \ \ \ \ $(3)

When both densities are of the identical value

$\left| \varepsilon _{g}\right| =\varepsilon _{ENU}\qquad \qquad \qquad
\qquad \qquad \qquad \qquad \qquad \ \ \ \ \ $(4)

then the following relations hold

$R=r_{ef}=(R_{g}.a)^{1/2}\qquad \qquad \qquad \qquad \qquad \qquad \ \ \ \ \
\ $(5)

in which $r_{ef}$ is the effective radius of a body with the mass {\it m},
$%
R_{g}$ is its gravitational diameter.

Compton wave $\lambda _{C}$ can be expressed as

$\lambda _{C}=\frac{\text{{\it
h\hskip-.2em\llap{\protect\rule[1.1ex]{.325em}{.1ex}}\hskip.2em%
}}}{m.c}\qquad \qquad \qquad \qquad \qquad \qquad \qquad \qquad \ \ \ \ \ \
\ $(6)

and then, based on identity of (5) and (6), relation expressing the lightest
particle capable to have gravitational influence on its environment appears
[3]

$m_{x}^{3}=\frac{\text{{\it
h\hskip-.2em\llap{\protect\rule[1.1ex]{.325em}{.1ex}}\hskip.2em%
}}^{2}}{2G.a}\qquad \qquad \qquad \qquad \qquad \qquad \qquad \qquad \ \ \ \
\ \ $(7)

The above mentioned relations are taken as starting points to manifest
mutual compatibility of the Vaidya metrics incorporating ENU model and
Newton gravitational theory.

In the Newtonian approach, relation describing the density of gravitational
energy is usually written as follows [2]

$\varepsilon _{g}=\frac{3G.m^{2}}{4\pi .r^{4}}\qquad \qquad \qquad \qquad
\qquad \qquad \qquad \qquad \ \ \ \ \ $(8)

Providing that

$\lambda _{C}=r\qquad \qquad \qquad \qquad \qquad \qquad \qquad \qquad
\qquad \ \ \ \ $(9)

the identity of (3) and (8) leads to (7) which means that the ENU model,
Vaidya metrics and Newton gravitational theory are mutually compatible.

Compatibility of the ENU model and the general theory of relativity will be
exemplified on Hawking model of black hole evaporation [4]. It was Hawking
who, stemming from principles of quantum mechanics, thermodynamics and
cosmology, evidenced that the output {\it P} of quantum evaporation of a
black hole with the diamater $R_{BH}$ is

$P=\frac{\text{{\it
h\hskip-.2em\llap{\protect\rule[1.1ex]{.325em}{.1ex}}\hskip.2em%
}}.c^{2}}{R_{BH}^{2}}$ \qquad \qquad \qquad \qquad \qquad \qquad \qquad
\qquad\ \ \ \ \ \ \ \ (10)

and the energy {\it E} of a single quantum of the evaporation is
\begin{verbatim}

\end{verbatim}

$E=\frac{\text{{\it
h\hskip-.2em\llap{\protect\rule[1.1ex]{.325em}{.1ex}}\hskip.2em%
}}.c}{R_{BH}}$ \qquad \qquad \qquad \qquad \qquad \qquad \qquad \qquad\ \ \
\ \ \ \ \ (11)

The mass $m_{\lim }$ of a limiting black hole is given [2, 4] by relation

$m_{\lim }=\left( \frac{\text{{\it
h\hskip-.2em\llap{\protect\rule[1.1ex]{.325em}{.1ex}}\hskip.2em%
}}.c^{4}.t}{4G^{2}}\right) ^{1/3}$\qquad \qquad \qquad \qquad \qquad \qquad\
\ \ \ \ \ \ \ \ (12)

where {\it t} is the cosmologic time. Comparing (11) and (12), for the
energy

of a limiting black hole evaporation quantum it follows

$E=m_{x}.c^{2}\qquad \qquad \qquad \qquad \qquad \qquad \qquad \qquad \ \ \
\ \ \ $(13)

where

$m_{x}=\left( \frac{\text{{\it
h\hskip-.2em\llap{\protect\rule[1.1ex]{.325em}{.1ex}}\hskip.2em%
}}^{2}}{2G.a}\right) ^{1/3}\qquad \qquad \qquad \qquad \qquad \qquad \qquad
\ \ \ \ $(14)

i.e. identity of (7) and (13) is manifested.

The mentioned identity can be taken as an example documenting compatibility
of the ENU model, Vaidya metrics and the general theory of relativity.

{\bf Conclusions}

The present results document the justifiability of Vaidya metrics
incorporation into mathematical tools used in cosmological problems solving;

The inclusion of Vaidya metrics in the ENU model allows to localize and
quantify the energy of gravitational field outside a body;

The present contribution suggests the existence of unity of the ENU model,
Newton gravitational theory and general theory of relativity.

{\bf References}

{\small 1. P.C. Vaidya, Proc. Indian Acad. Sci., {\bf A33} (1951) 264}

{\small 2. J. Sima, M. Sukenik, Preprint: gr-qc 9903090 (1999)}

{\small 3. J. Sima, M. Sukenik, M. Sukenikova, Preprint: qr-qc 9910094
(1999)%
}

{\small 4. S. Hawking, Sci. Amer.,{\bf \ 236} (1980) 34}

\end{document}